\documentclass[twocolumn,english,aps,pre,groupedaddress,showpacs]{revtex4}
\usepackage[T1]{fontenc}
\usepackage[latin1]{inputenc}
\usepackage{amsmath}
\usepackage{graphicx}
\usepackage{amssymb}

\makeatletter

\providecommand{\LyX}{L\kern-.1667em\lower.25em\hbox{Y}\kern-.125emX\@}

\usepackage[T1]{fontenc}
\usepackage[latin1]{inputenc}

\makeatletter

\usepackage[T1]{fontenc}
\usepackage[latin1]{inputenc}

\makeatletter

\usepackage[T1]{fontenc}
\usepackage[latin1]{inputenc}

\makeatletter

\usepackage[T1]{fontenc}
\usepackage[latin1]{inputenc}

\makeatletter

\usepackage[T1]{fontenc}
\usepackage[latin1]{inputenc}

\makeatletter

\usepackage[T1]{fontenc}
\usepackage[latin1]{inputenc}

\makeatletter

\usepackage[T1]{fontenc}
\usepackage[latin1]{inputenc}

\makeatletter

\usepackage[T1]{fontenc}
\usepackage[latin1]{inputenc}

\makeatletter

\usepackage{natbib}
\usepackage{graphics}
\usepackage{graphicx}
\usepackage{subfigure}
\usepackage{bm}
\usepackage{amsfonts}
\usepackage{amssymb}
\usepackage{amsmath}
\newcommand{\be}{\begin{equation}} 
\newcommand{\ee}{\end{equation}} 
\newcommand{\bea}{\begin{eqnarray}} 
\newcommand{\eea}{\end{eqnarray}}

\def\cd{\! \cdot \!}
\def\ah{{\bm H}}
\def\si{{\bm \sigma}}
\def\er{{\bm r}}
\def\en{{\bm n}}
\def\titah{{\bm h}}

\def\iks{{\bm x}}

\makeatother
\makeatother
\makeatother
\makeatother
\makeatother
\makeatother

\makeatother
\makeatother

\usepackage{babel}
\makeatother
\begin{document}
\title{Interaction model for magnetic holes in a ferrofluid layer}

\author{Renaud Toussaint}

\affiliation{Department of Physics, NTNU, N-7491 Trondheim, Norway}

\email{Renaud.Toussaint@fys.uio.no}

\author{J{\o}rgen Akselvoll}

\affiliation{Institutt for Energiteknikk, N-2007 Kjeller, Norway}

\author{Eirik G. Flekk{\o}y}

\affiliation{Department of Physics, University of Oslo, N-0316 Oslo, Norway}

\author{Geir Helgesen}

\affiliation{Institutt for Energiteknikk, N-2007 Kjeller, Norway}

\author{Arne T. Skjeltorp}

\affiliation{Institutt for Energiteknikk, N-2007 Kjeller, Norway}

\date{\today}

\begin{abstract} Nonmagnetic spheres confined in a ferrofluid layer
(magnetic holes) present dipolar interactions when an external magnetic
field is exerted. The interaction potential of a microsphere pair
is derived analytically, with a precise care for the boundary conditions
along the glass plates confining the system. Considering external
fields consisting of a constant normal component and a high frequency
rotating in-plane component, this interaction potential is averaged
over time to exhibit the average interparticular forces acting when
the imposed frequency exceeds the inverse of the viscous relaxation
time of the system. The existence of an equilibrium configuration
without contact between the particles is demonstrated for a whole
range of exciting fields, and the equilibrium separation distance
depending on the structure of the external field is established. The
stability of the system under out-of-plane buckling is also studied.
The dynamics of such a particle pair is simulated and validated by
experiments. 

\end{abstract} 



\maketitle

\section{Introduction}

The dynamic properties of so-called magnetic holes in ferrofluid layers
has been the object of increasing interest over the past twenty years
\cite{Skj83,HS91,Skj83b,Skj84b,Skj84,Skj95,Skj85,SH91,HS91c,HPS90,HPS90b,HS91b,MRu96,SCH01,SCH99,CHS98,CHS98b,PCH+96,HSM+88,WH85,DPM+86,SBe89,MRu95}.
These systems consist of spherical non-magnetic particles in a carrier
ferrofluid, whose size is order of magnitudes (1-100 $\mu m$) above
the one of the magnetic particles (0.01$\mu m$) in suspension. The
ferrofluid appears then as homogeneous at the scale of the large particles
-- holes --, and their effect on the magnetic field can be modelled
as a dipolar perturbation, where the magnetic moment is opposite to
the one of the displaced ferrofluid \cite{Skj83}. The system is generally
confined between glass plates in quasi-two-dimensional layers, whose
thickness slightly exceeds the diameter of the holes. The induced
dipolar interactions give rise to a rich zoology of physical phenomena,
such as crystallization of magnetic holes in constant or oscillating
magnetic fields \cite{Skj83,HS91,Skj83b,Skj84b,Skj84}, order-disorder
transitions in those crystals \cite{Skj95,Skj85,SH91,HS91c}, or non-linear
phenomena in the dynamics of those systems in low frequency oscillating
fields \cite{HPS90,HPS90b,HS91b,MRu96}, commonly described using
braid theory \cite{SCH01,SCH99,CHS98,CHS98b,PCH+96}.

The understanding of these systems is important in relation to industrial
ferrofluid applications \cite{CWo83,Ros87}, or for their potential
use in biomedicine \cite{USK+93,BEN+89,HPC+89}. The dynamics of these
phenomena can also be used indirectly to characterize the ferrofluid's
transport properties \cite{RuM93}, as its viscosity. Eventually,
the ability to shape the effective pair interaction potentials through
the imposed external magnetic field, makes these system good candidates
as large analog models to study phase transitions \cite{Skj87}, aggregation
phenomena \cite{HSM+88} or fracture phenomena in coupled granular/fluid
systems.

Nonetheless, despite the theoretical studies on similar dual systems
as ferromagnetic particles in a viscous fluid \cite{Lau51,NYa68},
and the extensive experimental observations of these magnetic holes,
there is a lack of theory describing this detailed effective pair
interaction potentials. Notably, there has been no satisfying explanation
so far for the existence of stable configurations of particles populations
with finite separation distances in external fields consisting of
a circular rotating inplane component and a constant normal one (reported
in \cite{HS91}), or for the existence of out-of-plane buckled structures
\cite{SH91}, and no theoretical framework for the influence of the
ferrofluid layer thickness (separation of the embedding plates).

We will show here how the magnetic boundary conditions along the confining
plates lead to rich effective interaction potentials rendering for
those structures, rather than the qualitative magneto-hydrodynamic
effects proposed in \cite{HS91}. In particular, we give an explanation
for the existence of a finite equilibrium separation between particles.
Theoretical work has already been done along this line \cite{WH85},
but in a reduced case of constant normal field. The present study
includes a circular high frequency oscillating field in addition.
The potential derived should be an essential brick in all the applications
mentioned above of the magnetic holes, and in general this type of
contribution of the confining structure should be relevant to any
quasi-2D colloidal system with a significant dielectric or magnetic
permeability contrast between the fluid medium and the confining structure,
as in \cite{MKF94}.

In this paper, we first describe the system under study and review
briefly the basic modelling asumptions and standard theory. We next
derive the instantaneous pair interaction potential with a precise
care for the magnetic permeability contrast of the boundaries, and
average it over the short oscillation periods to get the effective
interactions. We turn then to the properties of the equilibrium configurations,
sketch a simple dynamical theory and compare the theoretical and experimental
results for the dynamics of a particle pair. Eventually, the three
dimensional aspect of the ferrofluid layers is taken into account,
to evaluate both gravity-induced corrections, and the stability of
the system under out-of-plane buckling.

\section{System under study and basic assumptions }

\label{sect:description}

The system considered consists of two non-magnetic spheres inside
a ferrofluid which is homogeneous at their scale, whose susceptibility
and magnetic permeability are denoted respectively $\chi _{f}$ and
$\mu _{f}=\mu _{0}(1+\chi _{f})$, where $\mu _{0}=4\pi \cdot 10^{-7}H.m^{-1}$.
This ferrofuid is itself embedded between two glass plates considered
as perfectly plane, parallel and nonmagnetic. The magnetic field anywhere
between the glass plates is then decomposed between a uniform order
zero component resulting from the outer imposed field, plus a perturbation
due to the spheres. This perturbation is essentially dipolar: an isolated
sphere in a field with a given uniform (constant) boundary value $\ah $
at infinity provokes a purely dipolar perturbation outside it, generated
by an effective dipole as shown in Figure \ref{Fig:hole,pair}: \begin{figure} {\par\centering \resizebox*{0.35\textwidth}{0.07\textheight}{\includegraphics{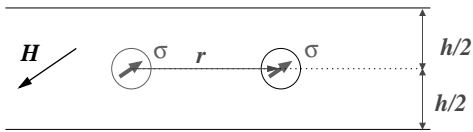}} \par}

\caption{Pair of non magnetic particles in a ferrofluid layer and related effective dipolar moments due to the external  magnetic field.\label{Fig:hole,pair}} \end{figure} \begin{eqnarray}
\si  & = & -V\bar{\chi }\ah \label{eq:effective,dipole}\\
\bar{\chi } & = & \frac{\chi _{f}}{1+2\chi _{f}/3}\label{eq:effective,suscept}
\end{eqnarray}
 where $\bar{\chi }$ is an effective susceptibility including a demagnetization
factor, i.e. whose precise form results when the boundary conditions
for the magnetic field on the surface of the sphere are properly taken
into account (see for example \cite{WH85,BB78}) -- SI magnetic units
are adopted throughout this whole paper, $V=\pi a^{3}/6$ and $a$
refer respectively to the volume and diameter of a sphere. This result
justifys the name of magnetic holes used for those nonmagnetic spheres,
and stays valid to leading order when a system more complicated than
an isolated sphere is considered: it holds as soon as the external
magnetic field $\ah $ can be considered as uniform at the scale of
a sphere, which will be assumed and commented further in section \ref{sub:Experimental-results-and}.

The average magnetic field inside the ferrofluid $\ah $ itself is
simply related to the external uniform magnetic field imposed outside
the glass plates $\ah ^{e}$ through \begin{equation}
\ah =\ah _{\Vert }^{e}+\frac{1}{1+\chi _{f}}\ah _{\bot }^{e}\label{eq:rel,H,dedans,dehors}\end{equation}
 to fulfill the Boundary Conditions along the glass-ferrofluid planar
interfaces -- namely, $\ah _{\Vert }^{e}=\ah _{\Vert }$ and ${\bm B}_{\perp }^{e}={\bm B}_{\perp }$
\cite{BB78,Kno00} -- where the parallel and normal components are
oriented relative to the glass plates.

The simplest model for a particle pair can be obtained first by neglecting
the effect of the nonmagnetic boundaries: considering a couple of
two such spheres of identical moments $\si $ with a separation vector
from center to center $\er $, the total interaction energy of the
system is, after Bleaney and Bleaney \cite{BB78}, \begin{equation}
U=\frac{\mu _{f}}{4\pi }\si ^{2}\left(\frac{1-3\cos ^{2}\theta }{r^{3}}\right)\label{eq:U,inst,zero}\end{equation}
 where $r=\left\Vert \er \right\Vert $ and $\theta =(\widehat{\er ;\si })$
is the angle between the field and the separation vector.

We consider now external fields composed of a circular in-plane component
oscillating at frequency $\nu $, superimposed to a constant normal
one. At high enough frequency, the relative displacement of the particles
during an oscillation period is negligible compared to its average
value, and the time dependence of the separation vector can be decoupled
in a slow and a fast varying mode, $\mathbf{r}(t)=\bar{\mathbf{r}}(t)+\delta \mathbf{r}(\bar{\mathbf{r}};t)$,
with $ \overline{\bf \delta r}=0 $ and $\delta r\ll \bar{r}$ . Throughout
the remainder of this paper , bold symbols refer to vectors, normal
ones to their norm, and the upper bar refers to averaged quantities
over oscillating time. Moreover, we will now use the additional constraint
on $\er $ that it should be in-plane, under the effect of the glass
plates over the microspheres which center them at positions equally
separated from the top and bottom boundaries. The presence of the
boundaries can indeed be represented by equivalent image dipoles outside
the confining plates, as established in section \ref{sect:images},
which repel the dipole from the boundaries. This constraint will be
adressed specifically in section \ref{sect:corrections}, to show
that this centering magnetic effect is valid in most work cases when
no lateral confinement is exerted in the system -- section \ref{sect:out-of-plane}
. The possible sink caused by gravity under the density contrast between
the microspheres and the fluid is generally negligible -- section
\ref{ssect:gravit,effects}. Thus, in the reference frame $(\hat{\bar{\mathbf{r}}};\hat{\mathbf{n}}\otimes \hat{\bar{\mathbf{r}}};\hat{\mathbf{n}})$
-- where hats refer to unit vectors, instantaneous fields read \begin{equation}
\ah =H(\cos \phi \sin \alpha ;\sin \phi \sin \alpha ,\cos \alpha )^{T}\label{eq:coord,h}\end{equation}
with by definition \begin{equation}
\text {cotan}\alpha =\beta =H_{\perp }/H_{\parallel }=H_{\perp }^{e}/(1+\chi _{f})H_{\parallel }^{e},\label{eq:def,beta}\end{equation}
 and $\phi =2\pi \nu t$. Thus, defining $\gamma =(\widehat{\bar{\mathbf{r}};\er })$,
comes $\cos ^{2}\theta =\cos ^{2}(\phi -\gamma )\sin ^{2}\alpha =\cos ^{2}(\phi -\gamma )/(1+\beta ^{2}),$
and \begin{eqnarray}
U(\er ;\phi ) & = & A\frac{a^{3}}{r^{3}}\left(2\beta ^{2}-1-3\cos (2\phi -2\gamma )\right)\label{eq:U,inst,simple}\\
\text {with\, }A & = & \frac{\mu _{f}\sigma ^{2}}{8\pi a^{3}(1+\beta ^{2})}=\frac{\mu _{f}\pi a^{3}\chi ^{2}H_{\parallel }^{2}}{288}.\label{eq:mult,fact}
\end{eqnarray}
 Consider distant enough particles to prevent contact and significant
hydrodynamic interactions during an oscillation period of the external
field: the neglect of inertial terms allows to balance magnetic interaction
forces and Stokes drag for both particles, which gives to leading
order in $\delta r/\bar{r}$,\begin{eqnarray}
\lefteqn{3\pi \eta a\cdot (\dot{\bar{\mathbf{r}}}+\delta \dot{\mathbf{r}})=-\nabla U(\er )} &  & \nonumber \\
 & \! \! = & \! \! \frac{3Aa^{3}}{\bar{r}^{4}}[(2\beta ^{2}-1)\hat{\bar{\mathbf{r}}}+3\cos (2\phi )\hat{\bar{\mathbf{r}}}+2\sin (2\phi )\hat{\mathbf{n}}\otimes \hat{\bar{\mathbf{r}}}]\label{eq:oscillating,pot}
\end{eqnarray}
Neglecting inertial terms is easily justified since a large upper
bound of Reynold's number can be evaluated as being $Re=\rho xa^{2}\nu /\eta \leq 10^{-4}$
where $x=\delta r/a$ is the relative amplitude of the oscillations
which will be shown straightforwardly to be typically below $10^{-2}$,
for typical diameters $a=50\mu m$, ferrofluid's viscosity $\eta =9\cdot 10^{-3}\, Pa.s$
and density $\rho =1000\, kg.m^{-3}$, and field oscillation frequency
$\nu =100\, Hz$. Random thermal motion in the ferrofluid is essentially
irrelevant at these size scales, as will be shown in section \ref{sub:Experimental-results-and},
justifying the use of deterministic dynamics instead of a Brownian
one. The above Eqs. (\ref{eq:U,inst,simple};\ref{eq:oscillating,pot})
establish that the slow motion $\bar{\mathbf{r}}$ is driven by an
effective potential obtained through time-averaging over the oscillations
of the field, i.e. simply using $\overline{\cos ^{2}\phi }=1/2$ at
fixed $\mathbf{r}=\bar{\mathbf{r}}$, \begin{eqnarray}
3\pi \eta a\dot{\bar{\mathbf{r}}} & = & -\nabla \overline{U}(\bar{\mathbf{r}})\label{eq:slow,ev.}\\
\text {with\, }\overline{U}(\bar{\mathbf{r}}) & = & A\frac{a^{3}}{\bar{r}^{3}}(2\beta ^{2}-1)\label{eq:pot,av}
\end{eqnarray}
while small and quick elliptic oscillations are performed,\begin{eqnarray*}
\delta \mathbf{r} & = & -\frac{\nu _{c}}{\nu }\frac{a^{5}}{\bar{r}^{5}}[3\sin (4\pi \nu t)\bar{\mathbf{r}}+2\cos (4\pi \nu t)\hat{\mathbf{n}}\otimes \bar{\mathbf{r}}]\\
\text {with\, }\nu _{c} & = & \frac{A}{4\pi ^{2}\eta a^{3}}=\frac{\mu _{0}(1+\chi _{f})\bar{\chi }^{2}H_{\parallel }^{2}}{1152\pi \eta }
\end{eqnarray*}
The relative magnitude of the fast oscillations $\delta r/\bar{r}=\nu _{c}a^{5}/\nu \bar{r}^{5}$
are indeed negligible as soon $\nu \gg \nu _{c}\simeq 0.1\, Hz$ for
typical ferrofluids, $\chi _{f}=1.9$, $\overline{\chi }=0.84$, $\eta =9\cdot 10^{-3}Pa.s$
and fields $H_{\parallel }=14\, Oe$. $\nu _{c}$ , inverse of the
viscous relaxation time of the system, is the critical frequency introduced
in \cite{HPS90b}, above which a particle pair cannot anymore follow
the direction of an external rotating field due to the fluid drag.
In this paper, we study regimes where $\nu \geq 10\, Hz$, for which
the relative variations of the separation vector are below 1\%, and
focus on the slow motion of the particles $\bar{\mathbf{r}}(t)$ driven
by $\overline{U}(\bar{\mathbf{r}})$.

In this simple picture however, this average potential is a simple
central one whose inverse cubic range reflects the dipole-dipole nature
of the interactions, and in the absence of any characteristic length
scale, this basic model predicts a very simple behavior for the particle
pair: depending on the ratio $\beta $ of the normal over the in-plane
field, either the two particles will repel each other without end
if $\beta >\beta _{c}=1/\sqrt{2}$, or they will attract each other
when $\beta <\beta _{c}$, until the magnetic forces are balanced
by contact forces or very short-range hydrodynamic forces sensitive
when particles almost touch -- when $(r-a)/a$ gets insignificant:
this theory is obviously insufficient to render for the finite equilibrium
separation distance, sometimes at a few diameters, which is experimentally
observed for a whole range of imposed fields \cite{HS91}. A proper
treatment of the boundary conditions of the system along the glass
plates, introducing the plate separation as an extra length scale
to the problem, will in the next section be shown to remedy this problem.

\section{Effect of the boundaries on the instantaneous interactions}

\label{sect:images}

\begin{figure}
\par\centering \resizebox*{0.4\textwidth}{0.3\textheight}{\includegraphics{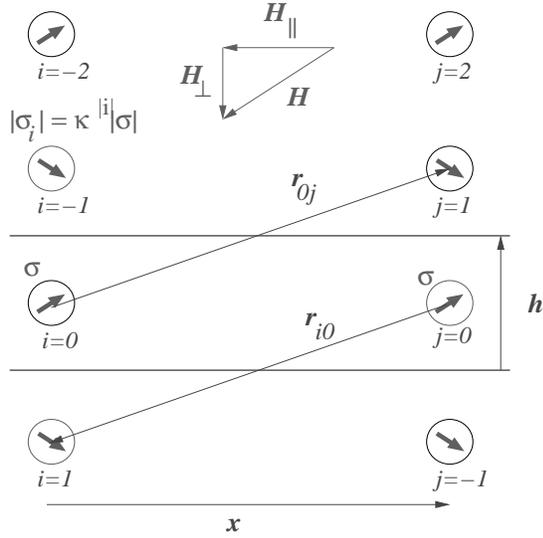}} \par{}

\caption{Image dipoles representing the Boundary Conditions.\label{fig:Images}}
\end{figure}

The boundaries between the ferrofluid and the embedding glass plates
are supposed to be perfectly plane. The two microspheres are supposed
perfectly centered between the glass plates, and the perturbation
of the magnetic field due to the presence of those spheres is
modeled as a perturbation due to two identical point-like dipoles
$\si =-\overline{\chi }V\ah $ located at the center of the spheres.
To fulfill the magnetic boundary conditions -- i.e. the continuity
of $H_{\Vert }$ and $B_{\perp }$-- along the plates, a direct use
of the image method (e.g. Weber \cite{Web50}) shows that this magnetic
perturbation between the plates is equal to the field emitted, in
an unbounded uniform medium of susceptibility $\chi _{f}$, by an
infinite series of dipoles: the two original ones, at locations defined
as $\bm 0$ and $\iks $, plus an infinite set of images for each
of them corresponding to the mirror symmetry across the plane boundaries
of the sources and all of the successive images. A magnification factor\begin{equation}
\kappa =\frac{\mu _{f}-\mu _{0}}{\mu _{f}+\mu _{0}}=\frac{\chi _{f}}{\chi _{f}+2}\label{eq:def,kappa}\end{equation}
 multiplies the amplitude of the dipolar moments at each symmetry
operation, i.e. explicitly defined by the following conditions:

For any dipole source $\si $ at position $\er _{0}$, with $\bm h$
the normal separation vector between the plates, an infinite set of
dipolar images indexed by $l\in \mathbb{Z}$ is defined by their locations
and moments -- see figure \ref{fig:Images}: \begin{eqnarray}
\er _{l}-\er _{0} & = & l\titah \label{eq:im,pos}\\
\si _{l_{\Vert }} & = & \kappa ^{|l|}\si _{\Vert }\label{eq:im,mom,inp}\\
\si _{l_{\perp }} & = & \kappa ^{|l|}(-1)^{\left|l\right|}\si _{\perp }\label{eq:im,mom,norm}
\end{eqnarray}

The total interaction energy of such a system -- where all dipoles
do not have anymore the same moment -- is (Bleaney and Bleaney \cite{BB78})
\begin{equation}
U=\frac{\mu _{f}}{8\pi }\sum _{b\neq c}\frac{1}{r_{bc}^{3}}\left[\si _{b}\cd \si _{c}-\frac{3\left(\si _{b}\cd \er _{bc}\right)\left(\si _{c}\cd \er _{bc}\right)}{\er _{bc}^{2}}\right]\label{eq:U,int,inst,im}\end{equation}
 where the sum on the $b$ index runs over the two source dipoles,
and the one over the $c$ index runs over the whole set of sources
and image dipoles -- as seen in the preceding section, separation
vectors can be considered as constant over the quick variations of
$\si $, and the upper bars over $\er $ are implicit for the remainder.

We will then use the following straightforward geometrical equalities
resulting from Eqs. (\ref{eq:effective,dipole},\ref{eq:def,beta},\ref{eq:im,pos}-\ref{eq:im,mom,norm})
: if the $c$ index represents the $l$-th image of the source $b$,
then \begin{eqnarray}
\frac{\si _{b}\cd \si _{c}}{\kappa ^{|l|}\sigma ^{2}} & = & \frac{1+(-1)^{l}\beta ^{2}}{1+\beta ^{2}}\label{eq:geom1}\\
\frac{\left(\si _{b}\cd \er _{bc}\right)\left(\si _{c}\cd \er _{bc}\right)}{\kappa ^{|l|}\sigma ^{2}\er _{bc}^{2}} & = & \frac{\beta ^{2}(-1)^{l}}{1+\beta ^{2}}\label{eq:geom2}\\
r_{bc} & = & |l|h.\label{eq:geomR1}
\end{eqnarray}
If on contrary $c$ represents the $l$-th image of one source --
as indexed in Figure \ref{fig:Images} -- , and $b$ the other source,
then \begin{eqnarray}
\frac{\si _{b}\cd \si _{c}}{\kappa ^{|l|}\sigma ^{2}}\! \!  & \! \! =\! \!  & \! \! \frac{1+(-1)^{l}\beta ^{2}}{1+\beta ^{2}}\label{eq:geom3}\\
\frac{\left(\si _{b}\! \cd \er _{bc}\right)\! \! \left(\si _{c}\! \cd \er _{bc}\right)}{\kappa ^{|l|}\sigma ^{2}\er _{bc}^{2}}\! \!  & \! \! =\! \!  & \! \! \frac{\left(y\cos \phi \! +\! \! \beta (-1)^{l}l\right)\! \! \left(y\cos \phi \! +\! \! \beta l\right)}{\left(1+\beta ^{2}\right)\left(y^{2}+l^{2}\right)}\label{eq:geom4}\\
r_{bc}\! \!  & \! \! =\! \!  & \! \! \sqrt{x^{2}+l^{2}h^{2}}\label{eq:geomR2}
\end{eqnarray}
 where \begin{equation}
y=\frac{x}{h}\label{eq:def,y,ratio,separations}\end{equation}
 is the ratio of the particle in-plane separation distance to the
plate separation. Introducing those equalities in Eq. (\ref{eq:U,int,inst,im})
we get: \begin{eqnarray}
\! \! \! \! \! \frac{U}{2Aa^{3}} & = & \frac{I_{0}^{so}}{x^{3}}+\sum _{l\in \mathbb{Z}^{*}}\kappa ^{|l|}\left[\frac{I_{l}^{ss}}{|l|^{3}h^{3}}+\frac{I_{l}^{so}}{(x^{2}+l^{2}h^{2})^{3/2}}\right]\label{eq:U,int,inst,im,bis}\\
\! \! \! \! \! I_{l}^{ss} & = & 1-2(-1)^{|l|}\beta ^{2}\label{eq:int,own,im,inst}\\
\! \! \! \! \! I_{l}^{so} & = & \left[1+(-1)^{|l|}\beta ^{2}\right.\nonumber \\
 &  & \left.-3\frac{\left(y\cos \phi \! +\! \! \beta (-1)^{|l|}l\right)\! \! \left(y\cos \phi \! +\! \! \beta l\right)}{\left(y^{2}+l^{2}\right)}\right]\label{eq:int,other,im,inst}
\end{eqnarray}
 where $A$ is the constant defined in Eq. (\ref{eq:mult,fact}).
The first term, denoted by $I_{0}^{ss}$, is the direct interaction
between the two sources, the next one $I_{l}^{ss}$ corresponds to
the interactions between the sources and their own images, and finally
the term $I_{l}^{so}$ corresponds to the cross-interactions between
a source and the images of the other one.

\section{Time-averaged effective interactions}

\label{sect:isotropic}

\subsection{Derivation of the potential}

\label{ssect:estab,isot}

The $\ah $-fields consist of a perfectly circular in-plane component
$\ah _{\Vert }=H_{\Vert }(\cos \phi \hat{\er }+\sin \phi \hat{\en }\wedge \hat{\er })$,
and a normal component $\ah _{\perp }$ which is maintained constant.
During an oscillation of the in-plane field, the only significantly
varying quantity in Eq. (\ref{eq:U,int,inst,im}) is the angle $\phi $,
with once again $\overline{\cos ^{2}\phi }=1/2$.

The term $I_{l}^{ss}$, e.g. the interactions between a source and
its own images, naturally does not depend on the in-plane separation
vector $\iks $, and produces no net in-plane force. It is worthwhile
to note that this would however not be the case for any non-plane
interfaces, giving the possibility to quench any geometrical property
of the roughness of the interfaces in this potential. In the current
hypothesis of purely planar plates, this interaction term is only
responsible for a normal centering force, as will be established in
section \ref{sect:out-of-plane}. For the present purpose where the
microspheres are constrained on the half-plane between the plates,
the in-plane forces are the only relevant ones, and this term is simply
discarded in the following.

The remaining terms produce through time-average the interaction energy\begin{equation}
\overline{U}(\iks )=A\frac{d^{3}}{h^{3}}u\left(\frac{x}{h}\right)\label{eq:U,av,isot,with,dim}\end{equation}
 with $A$ the constant defined in Eq. (\ref{eq:mult,fact}), and
a dimensionless term

\begin{eqnarray}
u(y)\! \!  & \! \! =\! \!  & \! \! (2\beta ^{2}-1)y^{-3}\label{eq:U,av,im,isot}\\
 &  & +4\sum _{l=1}^{+\infty }\kappa ^{l}\left(\frac{1+(-1)^{l}\beta ^{2}}{(y^{2}+l^{2})^{3/2}}-\frac{3}{2}\frac{y^{2}+2(-1)^{l}l^{2}\beta ^{2}}{(y^{2}+l^{2})^{5/2}}\right)\nonumber 
\end{eqnarray}
 The first term reduces to the expression of the first-order theory
of Eq. (\ref{eq:pot,av}), which is a test of self-consistency, since
an infinite medium would be equivalent to the absence of permeability
contrast along the plates, $\kappa =0$. The following ones render
for the interactions between a source and the images of the other
one. Due to the cylindrical symmetry of the problem with a purely
circular in-plane field, this interaction is isotropic: the dependence
on the separation vector $\iks $ enters only through its norm $x=hy$.

\subsection{Properties of the isotropic interactions}

\label{ssect:propert,isot}

The introduction of those images is responsible for possible finite
separation equilibrium distances for a whole range of $\beta $ characterizing
the imposed magnetic field, as is illustrated in Figure \ref{fig:interaction,typolog}:
\begin{figure} {\par\centering \subfigure[Purely attractive, $\beta < \beta_m$]{\resizebox*{0.4\textwidth}{0.17\textheight}{\includegraphics{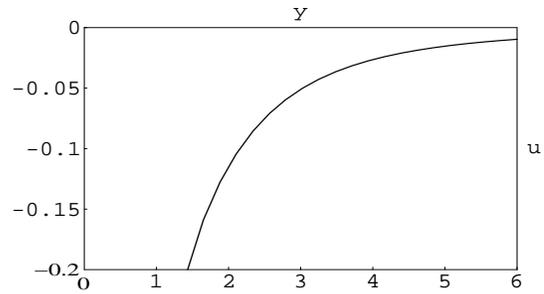}}} \subfigure[Metastable, $\beta_m < \beta < \beta_c$]{\resizebox*{0.4\textwidth}{0.185\textheight}{\includegraphics{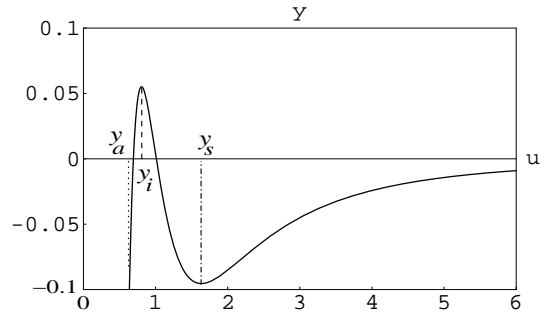}}} \par}

{\par\centering \subfigure[Stable, \quad $\beta_c < \beta < \beta_u$]{\resizebox*{0.4\textwidth}{0.183\textheight}{\includegraphics{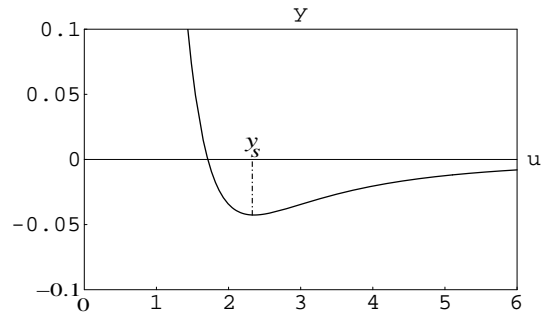}}} \subfigure[Purely repulsive, $\beta_u < \beta $]{\resizebox*{0.4\textwidth}{0.17\textheight}{\includegraphics{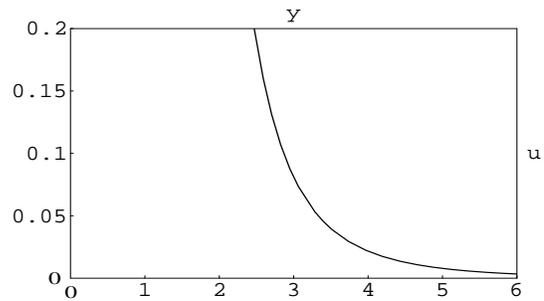}}} \par}

\caption{Possible types of the interactions depending on \protect\( \beta \protect \). \label{fig:interaction,typolog}} \end{figure}In this example, a typical susceptibility $\chi _{f}=1.9$ was considered
for the ferrofluid, i.e. $\kappa =0.49$. The four potentials $u(y)$
represented correspond respectively to $\beta =0.4,\, 0.58,\, 0.8,\, 2.1$.
They were obtained by truncating the sums in Eq. (\ref{eq:U,av,im,isot})
to order $l=10$, which corresponds to a relative error lower than
$10^{-3}$ in $u$ for any $y$.

The general typology of those potentials can be classified in four
cases as will be demonstrated in the next section, separated by three
particular values of $\beta $, referred to as $\beta _{m}$, $\beta _{c}$
and $\beta_u$ -- which depend only on the susceptibility $\chi _{f}$:

\begin{itemize}
\item (a) For $\beta <\beta _{m}$ , $u$ is a monotonically increasing
function, and the magnetic forces are purely attractive, thus leading
any pair of spheres to contact.
\item (b) For $\beta _{m}<\beta <\beta _{c}$, $u$ presents a short-range
attractive core, and presents both a maximum -- an instable equilibrium
point -- at some distance $y_{i}$, typically slightly below $1$,
and a minimum -- a stable equibrium point -- at a greater distance
$y_{s}$, usually above $1$.\\
Two locally stable equilibrium configurations are possible in principle,
depending on the ratio $y_{a}=a/h$ of the particle diameter to the
plate separation, and of the initial particle separation $y_{\text {init}}$:
if $y_{\text {init}}>y_{i}$, the particles should end up at the equilibrium
separation $y_{s}$, which since $y_{s}>1>y_{a}$ corresponds to an
equilibrium configuration without contact between the particles. If
on contrary $y_{\text {init}}<y_{i}$, the particles should attract
each other and end up in contact at $y_{a}$. Since any separation
$y<y_{a}$ is forbidden due to contact forces, this second case is
only possible when $y_{a}<y_{i}$, i.e. when the ratio of plate separation
over particle diameter is sufficiently big. In that case, if the thermal
fluctuations are large enough to let the particles go over the energy
maximum at $y_{i}$ with a significant probability over the observation
time, only one of the two possible equilibrium configurations will
be thermodynamically stable, and the other one will be only metastable.
The selection of stability / metastability between the two is determinated
by the comparison of $u(y_{a})$ and $u(y_{s})$.
\item (c) For $\beta _{c}<\beta <\beta_u$, $u$ presents only a global
minimum, which thus corresponds to a stable equilibrium separation
distance $y_{s}$. Since generally $y_{s}>1>y_{a}$, this equilibrium
configuration corresponds to a finite separation distance $hy_{s}-a>0$.
\item (d) For $\beta_u<\beta $, $u$ is monotonically decreasing, and
the time averaged magnetic forces are purely repulsive at any separation.
\end{itemize}

\section{equilibrium separation distance as function of the applied field}

\label{sect:eq-sep-dist}

The separation equilibrium distance $y_{eq}$ -- which is stable or
not -- correspond to the extrema of the potential, and can be obtained
in principle by solving \begin{equation}
\frac{du}{dy}(y_{eq})=0.\label{eq:principle,eq,dist}\end{equation}
$y_{eq}$ corresponds to $y_{s}$ or $y_{i}$ defined in section \ref{ssect:propert,isot},
according to the sign of $d^{2}u/dy^{2}$. Derivating Eq. (\ref{eq:U,av,im,isot})
with respect to the scaled separation $y$, leads straightforwardly
to \begin{eqnarray}
\frac{y^{4}}{3}\! \frac{du}{dy}\!  & \! =\!  & \! (1-2\beta ^{2})\! -\! 2\sum _{l=1}^{+\infty }\! \kappa ^{l}\lambda _{l}(y)\! (-1+2(-1)^{l}\beta ^{2})\label{eq:der,u,y,isot}\\
\lambda _{l}(y)\!  & \! =\!  & \! \frac{(y^{2}-4l^{2})y^{5}}{(y^{2}+l^{2})^{7/2}}\label{eq:expr,lambda,l}
\end{eqnarray}
The Eq. (\ref{eq:der,u,y,isot}) above is the sum of a term independent
of $\beta $, plus another proportional to $\beta ^{2}$. The constant
term can be shown to be positive, and the prefactor of $\beta ^{2}$
strictly negative, for any possible $(y,\kappa )$ -- i.e. any $y>0,0\leq \kappa <1$.
Thus, $du/dy(y,\beta )$ is a monotonic decreasing function of $\beta ^{2}$,
equal to zero when\begin{equation}
\beta _{0}(y)=\sqrt{\frac{1+2\sum _{l=1}^{+\infty }\kappa ^{l}\lambda _{l}(y)}{2+4\sum _{l=1}^{+\infty }(-1)^{l}\kappa ^{l}\lambda _{l}(y)}}\label{eq:beta,funct,y,eq,isot}\end{equation}
 Thus, for a given field configuration $\beta $, and separation $y$,
pair interactions are attractive, i.e. $du/dy>0$, if $\beta <\beta _{0}(y)$,
and conversely if $\beta >\beta _{0}(y)$. A numerical study of the
above function, for any $\kappa $, shows that $\beta _{0}(y)$ is
monotonically decreasing from $\beta _{c}=1/\sqrt{2}$ to a finite
positive minimum $\beta _{m}(\kappa )$, between $y=0$ to $y_{m}(\kappa )$,
and next monotonically increasing up to a finite limit $\beta_u(\kappa )>\beta _{c}$
between $y_{m}(\kappa )$ and $y\rightarrow +\infty $. 

These considerations allow to obtain by a direct graphical inversion
of $\beta _{0}(y)$, the possible roots $y_{s}(\beta )$ and $y_{i}(\beta )$
for which the interaction forces are zero at a given field geometry
$\beta $, as shown in Figure \ref{fig:sep,eq,beta,iso} which is
obtained for the particular case $\chi _{f}=1.9,$ i.e. $\kappa =0.49$.\begin{figure} {\par\centering \resizebox*{0.45\textwidth}{0.3\textheight}{\includegraphics{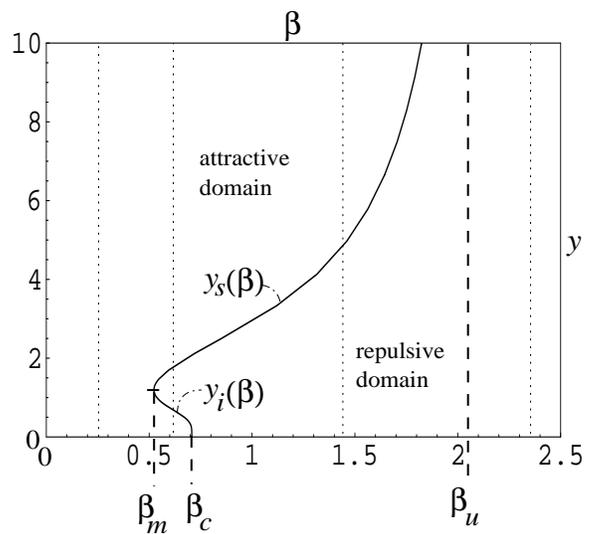}} \par}

\caption{Equilibrium separation of a pair as function of the applied field\label{fig:sep,eq,beta,iso}} \end{figure} This nonlinear dependence of the equilibrium separation $y_{s}(\beta )$
seems more complex than observed in earlier experiments by Helgesen
and Skjeltorp \cite{HS91}. This apparent discrepancy will be resolved
in the next section, which is centered on finite time results.

Identifying the three parameters $\beta _{m},\beta _{c},\beta_u$
identified above as \begin{eqnarray}
\beta _{c} & = & \beta _{0}(y=0)\label{eq:beta,c}\\
\beta _{m} & = & \min _{y}\beta _{0}(y)=\beta _{0}(y_{m})\label{eq:beta,m}\\
\beta_u & = & \lim _{y\rightarrow +\infty }\beta _{0}(y),\label{eq:beta,M}
\end{eqnarray}
 these arguments prove that the pair effective potentials belong to
one of the four types described in the previous section:

(a) If $\beta <\beta _{m}$, the potential is purely attractive at
any separation.

(b) If $\beta _{m}<\beta <\beta _{c}$, there are two roots to Eq.
(\ref{eq:principle,eq,dist}), denoted $y_{i}(\beta )$ and $y_{s}(\beta )$:
the potential is attractive below $y_{i}$ or above $y_{s}$, and
repulsive between both.

(c) If $\beta _{c}<\beta <\beta_u$, the potential presents a single
minimum at $y_{s}(\beta )$.

(d) If $\beta >\beta_u$. the potential is purely repulsive and
there is no equilibrium separation.

The definition of $\beta_u$ above -- Eq. (\ref{eq:beta,M}) --
shows also clearly that $\lim _{\beta \rightarrow \beta_u^{-}}(y_{s})=+\infty $:
in principle, it should be possible to drive a pair of microspheres
in an equilibrium configuration with any desirable separation distance.
Naturally, since the magnetic interactions decay rapidly with distance,
thermal processes or any kind of external perturbation in the fluid
flow, or default in the planarity of the plates, will be predominent
at large separations, where this theory will become inapplicable. 

The dependence of $\beta _{c},\beta_u,\beta _{m}$ on the susceptibility
of the ferrofluid (through the parameter $\kappa $) is as follows:
Replacing $\lambda _{l}(0)=0$ in Eq. (\ref{eq:beta,funct,y,eq,isot})
shows that \begin{equation}
\beta _{c}=1/\sqrt{2}\label{eq:beta,c,value}\end{equation}
 independently of $\kappa $. Similarly, since $\lim _{y\rightarrow +\infty }\lambda _{l}(y)=1$,
$\beta_u$ is easily summed as \begin{eqnarray}
\beta_u & = & \frac{1}{\sqrt{2}}\sqrt{\frac{1+2\sum _{l=1}^{+\infty }\kappa ^{l}}{1+2\sum _{l=1}^{+\infty }(-\kappa )^{l}}}\nonumber \\
 & = & \frac{1}{\sqrt{2}}\frac{1+\kappa }{1-\kappa }\label{eq:beta,M,value}
\end{eqnarray}
A numerical study of $\beta _{m}$ shows that it decreases monotonically
with $\kappa $, down to zero when $\kappa \rightarrow 1$.

This shows that the range of $\beta $ over which stable equilibrium
distances exist is larger when the susceptibility of the ferrofluid
is important ($\kappa $ increases with $\chi _{f}$, $\beta _{m}$
and $\beta_u$ are respectively decreasing and increasing with
$\kappa $), up to the limiting case of an infinitly susceptible ferrofluid
$\kappa \rightarrow 1$ ($\chi _{f}\gg 1$), for which $\beta _{m}\rightarrow 0,\, \beta_u\rightarrow +\infty $,
and there is a finite stable equilibrium separation for any ratio
$\beta $.

The other limiting case $\kappa \rightarrow 0$ is obtained directly
by discarding the sums in Eq. (\ref{eq:beta,funct,y,eq,isot}) - their
convergence to $0$ in that limit for any $y$ is straightforward
-. This shows that $\lim _{\kappa \rightarrow 0}\beta _{m}=\lim _{\kappa \rightarrow 0}\beta_u=\beta _{c}=1/\sqrt{2}$:
without any permeability contrast along the glass plates, no images
are felt and the first-order theory is recovered - the interactions
are simply purely attractive when $\beta <\beta _{c}$, and purely
repulsive when $\beta >\beta _{c}$, with the stable regimes $\beta \in [\beta _{m},\beta_u]$
disappearing.

An overall picture of $y_{eq}(\beta )$ for different values of $\kappa $
distributed regularly between $0$ and $0.9$ is given in figure (\ref{Fig:Eq,dist,tous,kappa}).\begin{figure} {\par\centering \resizebox*{0.45\textwidth}{0.25\textheight}{\includegraphics{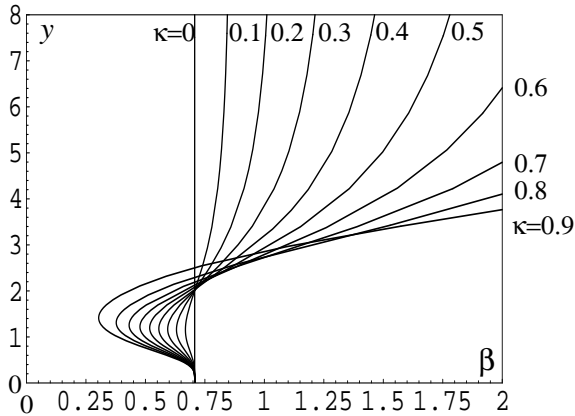}} \par}

\caption{Equilibrium separation distances as function of the applied field, for various ferrofluids.\label{Fig:Eq,dist,tous,kappa}} \end{figure} The sums in Eq. (\ref{eq:beta,funct,y,eq,isot}) have been truncated
to order 10, which results in an accuracy better than 1\% for the
displayed function $\beta _{0}(y)$ -- functions $\lambda _{l}(y)$
defined in Eq. (\ref{eq:expr,lambda,l}) can be bounded by $|\lambda _{l}(y)|<1/(y^{2}+l^{2})^{3/2}$,
so that $\left|\sum _{l=N}^{+\infty }\kappa ^{l}\lambda _{l}(y)\right|<\sum _{l=N}^{+\infty }\kappa ^{l}\left|\lambda _{l}(y)\right|<\kappa ^{N}/(1-\kappa )N^{3}$,
and similarly for $\left|\sum _{l=N}^{+\infty }(-\kappa )^{l}\lambda _{l}(y)\right|$.
Thus, neglecting terms from order $N=11$ in the sums, for any $\kappa <0.9$,
results in a relative error on $\beta _{0}(y)$, smaller than $0.9^{11}/(1-0.9)11^{3}\simeq 0.0023$.

\section{Finite-time theory and experiments\label{sec:Finite-time}}

\subsection{Simple time-dependent theory\label{sub:Simple-time-dependent-theory}}

The preceding section was centered on equilibrium properties of this
system, but the relaxation time to reach equilibrium amounts to hours
or days in certain configurations, as will be shown here. In order
to compare efficiently theoretical and experimental data, we have
therefore concentrated on the slow dynamics of this system, starting
from hole pairs in contact under the effect of a purely in-plane
field $\beta =0$, through the following scheme: neglecting once again
the inertial terms, Stokes drag and magnetic interactions are balanced
to obtain \begin{equation}
3\pi \eta a\frac{dx}{dt}=-\frac{d}{dx}\overline{U}(x)\label{eq:simple,slow,motion,model}\end{equation}
where the effective potential includes the images due to the boundaries,
Eq. (\ref{eq:U,av,isot,with,dim}). The viscosity $\eta $ above is
renormalized to take into account hydrodynamic interactions with the
confining plates, as will be detailed further. 

In dimensionless units, this equation leads to \begin{eqnarray}
T\frac{dy}{dt'} & = & -\frac{d}{dy}u(y)\label{eq:adimensioned,slow,motion}\\
\text {with\, }T & = & \frac{864\eta }{\mu _{0}(1+\chi _{f})\bar{\chi }^{2}H_{\parallel }^{2}}\frac{h^{5}}{d^{5}}\label{eq:characteristic,time}
\end{eqnarray}
and $y(t')=x/h,$ $t'=t/T$, and $u(y)$ the potential
defined in Eq. (\ref{eq:U,av,im,isot}). The characteristic time above
lies typically around $30\, s$ to $5\, min$ at usual working parameters,
as will be shown in the next section, and moreover the potential wells
can be pretty flat, thus producing often metastable situations that
last from minutes to hours (the driving force close to
equilibrium position is proportional to the distance to it, times
the second derivative of the potential in the well, and $u''(y_{s})\rightarrow 0$
when $\beta \rightarrow \beta_u^{-}$).

Starting from a pair configuration in contact, and setting at time
$0$ the field parameters ($\beta $ ratio and magnitude $H_{\parallel }$)
to a constant value, the time $t'(y)$ to reach a given separation
will be directly obtained through numerical integration of the differential
Eq. (\ref{eq:adimensioned,slow,motion}):\begin{equation}
t'(y)=-T\int _{z=a/h}^{y}dz/u'(z).\label{eq:adimensioned,time,integrated}\end{equation}
An inverse representation of the above is plotted for a ferrofluid
of susceptibility $\chi _{f}=1.9$. In Figure \ref{Fig:adimen,time,sep,beta08},
the solid thin line represents $y(t')$ at fixed $\beta =0.8$, and
in Figure \ref{Fig:merged}, the dashed lines represent $y(\beta )$
for four characteristic times $t'=.25,\, 1,\, 4,\, 16$, slowly converging
to equilibrium separation $y_{s}(\beta )$, plotted as a solid line.

To compare this with experiments, the fine tuning of the time dependence
requires a refined analysis of the hydrodynamic interactions in the
above: since experiments are carried out in cells of width $h$ comparable
with the diameter $a$ of the embedded holes -- typically $h/a$ lies
between 1.1 and 2 --, a strong hydrodynamic coupling with the confining
plates is present. Considering that both particles sit at a fixed
fraction $z$ of the plate separation relative to the central position
(nonzero $z$ can be obtained in principle between horizontal plates
due to the density contrast between the particles and ferrofluid),
these interactions are represented according to \cite{FLi94}, by
a normalization of the Stokes drag as \begin{equation}
\eta /\eta _{0}=f[a/(h+2zh)]+f[a/(h-2zh)]-1\label{eq:renormalize,viscosity}\end{equation}
with $\eta _{0}$ the naked viscosity of the carrier ferrofluid, and\begin{equation}
f(x)=\left(1-\frac{9x}{16}+\frac{x^{3}}{8}-\frac{45x^{4}}{256}-\frac{x^{5}}{16}\right)^{-1}+O(x^{6}).\label{eq:detailed,renormalize,viscosity}\end{equation}
Two experimental cases will be considered here, where $h/a=1.4;\, z\simeq 0$
and $h/a=2,\, z\simeq 0.14$, corresponding respectively to $\eta /\eta _{0}\simeq 2.6$
and $2.4$. This is consistent with Faucheux and Libchaber's measures
for a confined brownian motion \cite{FLi94} , and with an experimental
value $\eta /\eta _{0}\simeq 2.4$, that we measured in the $h/a=1.4$
case by placing $50\, \mu m$ diameter particles between $70\, \mu m$
distant plates set up in vertical position, without any magnetic field,
and by recording the motion of a single particle under the effect of
buoyancy forces. The density of the ferrofluid was $\rho _{f}=1.24$,
the one of the polystyrene particles $\rho _{p}\simeq 1$. This value
is slightly below the theoretical $\eta /\eta _{0}\simeq 2.6$, which
is consistent with the effect of the brownian motion of the particle
along the normal direction, as analyzed in \cite{FLi94}. We have
chosen to use $\eta /\eta _{0}=2.5$ in the following. 

Hydrodynamic interactions between both particles were neglected,
which should be a relatively poor approximation for particles close
to contact, but become reasonable at separations $x/h>2$ where
most of the time is spent to achieve equilibrium at relatively large
separations, and is therefore the most important one in the present
context. The typical magnitude of this error can be roughly estimated
through the analyzes performed by Dufresne et al. or Grier and Behrens
\cite{DPM+86,GB01} for a close context: they derived for two particles
of diameter $a$, separated by $x$ and at a distance $h/2$ from
a single plate, the hydrodynamic corrections to the mobility to first
order in $x/a$ and $x/h$. Considering for simplicity a double contribution
for two plates relatively to Eq. (13) in \cite{DPM+86}, contributions
due to the particle-particle interactions, become for $h/a=1.4;\, x/h>2$,
less than 30\% of the one due to the plates -- term $9x/16$ in Eq.
(\ref{eq:detailed,renormalize,viscosity}). 

We neglect also the rotational degrees of freedom of the ferrofluid itself, which can lead to the rotation of the nonmagnetic spheres, and induce another type hydrodynamic interactions between pairs of spheres: in ferrofluids submitted to circular magnetic fields, the rotation of the magnetite particles induces asymmetric stresses in the fluid,
 which leads to a counter-rotation of the magnetic holes 
\cite{HS91,MRu95}. The mismatch between this rotational motion of the magnetic holes and the one of the magnetites
could in principle induce a vortex in the fluid flow around each of the holes,
which would lead to a net hydrodynamic torque over close pairs of holes. 
Nonetheless, with the ferrofluid and field frequency regime used here (<100 Hz), this rotational motion of the holes is so slow 
that it is hardly detectable with the field
 frequency used ($\nu_f<100 Hz$). Experimentally, with a similar ferrofluid (kerosene-based, $\chi=0.8$),
 the hole's frequency was bounded by $\nu_s<0.01 Hz$. This rotational motion was described 
theoretically in detail by Miguel and Rubi \cite{MRu95}.
Through this theory, for the ferrofluid and field frequencies used here, the frequency of the magnetic holes will straightforwardly 
be shown to be lower than $\nu_s<0.05 Hz$.
This justifies for the present study the neglect of these rotation-induced hydrodynamic interactions. 
These interactions would in any case lead to a purely rotational motion of the hole pair,
decoupled from the purely central forces induced by the magnetic interactions studied here. 
Experimentally, some very slow rotational motion of the hole pairs was indeed occasionally observed  when the particles were close (non periodic, angular velocity always lower than 0.001 Hz),
but no systematic trend for its direction or velocity was noted, i.e. this effect was beyond the experimental error.

To obtain the theoretical estimate of the hole's frequency above, the Langevin parameter of the ferrofluid, defined \cite{MRu95} as 
$ \mu = m_0 H / k_B T $ where  $m_0$ is the magnetic moment of the magnetites, is derived from the saturation magnetization of the 
ferrofluid and its susceptibility \cite{Lar99} as 
$ \mu = 3 H / \chi M_{\text{sat}}. $
For the ferrofluid used, $\chi=1.9$, $M_\text{sat}=200 G$, $H=14$ Oe, and $\mu=0.26$. 
This enables to express the ratio of the hole's frequency over the field's one, as \cite{MRu95}
\begin{equation}
\nu_s/\nu_f=- \Phi (\mu - \text{tanh} \mu)/ (\mu + \text{tanh} \mu) = - 0.00041
\end{equation}
where $\Phi=-0.036$ is the magnetite volume fraction of the ferrofluid.
For the regime  $\nu_f<100 Hz$ where the experiments are carried, the spheres frequency is thus below $\nu_s=0.041 Hz$.

\subsection{Experimental results and scaling\label{sub:Experimental-results-and}}

We used pairs of $a=50\, \mu m$ diameter neutral polystyrene spheres of density
$\rho _{p}=1$, designed according to Ugelstad's technique \cite{Uge80},
and ferrofluids of susceptibility $\chi _{f}=1.9$, density $\rho _{f}=1.24$
and viscosity $\eta _{0}=0.009\, Pa.s$ \cite{Fer}, confined in horizontal
cells of thickness $70\, \mu m$ or $100\, \mu m$ and width of order
centimeters. The thickness of the cell was obtained by confining plates by quenching a few 70 $\mu m$ diameter spheres between the two plates clamped together (these spacers where typically half a centimeter distant from each other). The oscillating inplane and constant normal field were
generated by external coils, with a typical magnitude $H_{\parallel }=H_{0}=14.2\, Oe$,
with frequencies from 10 to 100 $H\! z$. Magnitudes and phase of
the field are accurate up to 1\%, which also corresponds to the degree of homogeneity of the in and out-of-plane fields throughout the entire cell. The direction of the constant field varies slightly along the cell, with a maximum $3^{\circ }$ misalignment
from the direction normal to the confining planes (these accuracies for the homogeneity and misalignment of the field were obtained directly by considering the geometry of the Helmholtz coils generating the field, whose characteristic extent is 10 cm, together with a .5 mm accuracy for the position of the cell inside them). The in-plane motion of the particles was recorded using a microscope, and a camera linked
to a numerical data analysis setup. The entire experimental setup
is described, for example, in Ref. \cite{HS91b}. The samples were prepared
with a pure inplane field to bring the particles in contact, after
that the inplane field was maintained constant and the normal field
was set to a constant $H_{\perp }^{e}=(1+\chi _{f})\beta H_{\parallel }$.
The normal field required typically a few seconds to stabilize. Separation
between the particles were then recorded every $10\, s$, for $30\, min$.
The concentration of the spheres in the entire sample was such as the nearest sphere or spacer would sit at least 20 diameters apart from the observed pair,
 which was sufficiently dilute not to influnce the motion of the observed pair.

\subsubsection{Time-dependent result at fixed field geometry and discussion.}

The scaled separation as function of time is shown in Figure
\ref{Fig:adimen,time,sep,beta08}, for five experiments carried out
at $\beta =0.8$, i.e. for the potential represented in Figure \ref{fig:interaction,typolog} (c),
which presents a single minimum at $y_{s}(0.8)=2.35$. Four experiments
were carried out in a $h=70\, \mu m$ thick cell, two of them at identical
field amplitudes $H_{\parallel }=H_{0}=14.2\, Oe$, and two other
at $H_{\parallel }=0.7\, H_{0};\, 0.5\, H_{0}$. The last experiment
was carried out in a $h=100\, \mu m$ thick cell, with an inplane
field $H_{0}$. \begin{figure} {\par\centering \resizebox*{0.45\textwidth}{0.35\textheight}{\includegraphics{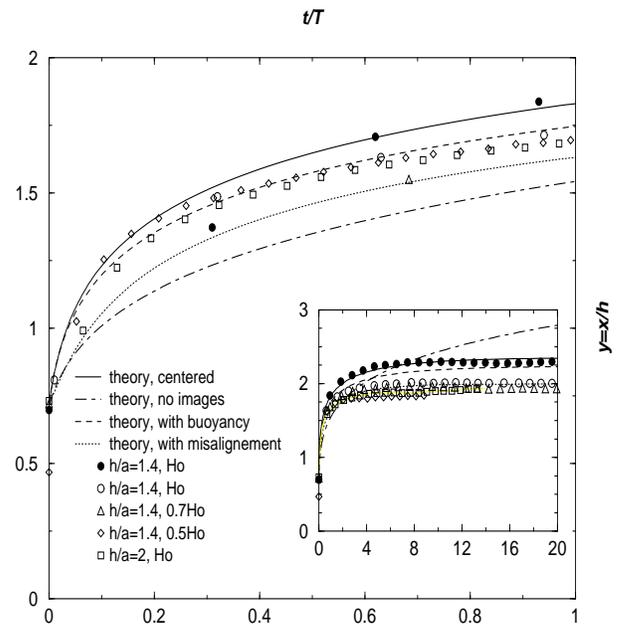}} \par}

\caption{Scaled separation as function of time for various field magnitudes and plate separations, for $\beta=H_{\perp}/H_{\parallel}=0.8$. The inset represents the same data for longer times.\label{Fig:adimen,time,sep,beta08}} \end{figure} These parameters corresponded respectively to characteristic times
evaluated using Eq. (\ref{eq:characteristic,time}) as $T=32\, s,\, 63\, s,\, 129s,\, 192s$.

\paragraph{Comparison with the theory.}
The large part of the figure represents the short time evolution $t<T$
(typically the first minutes), the encapsulated part shows longer
time (typically 30 min). Every data point (separated by 10 s) was
used for the early regime, one point out of 2 to 12 depending on the
experiment for the later one. At shorter times, the scaled data
collapse reasonably well on the theoretical curve obtained from the
theory sketched so far -- solid line. The only two main outliers (first
filled circle and triangle, $t=10s$) presumably corresponded to a
weaker normal field in the first few seconds, until it stabilized.
For illustration, the first-order theory neglecting the magnetic confinement
leads to the dot-dashed curve, obtained using Eq. (\ref{eq:adimensioned,time,integrated})
with a bare potential retaining only the source-source term. This
too simple theory clearly differs from data both at short and large
times. 

\paragraph{Corrections due to buoyancy.}
A third dashed theoretical curve was plotted, corresponding to a slightly
refined calculation of the interaction potential, where buoyancy forces
due to the density contrast between ferrofluid and particles led to
a shift $z$ of the particle pairs from the central mid-plane of the
cell. The relative displacement $z/h$ due to this gravitational correction
is evaluated in section \ref{ssect:gravit,effects} as $15\%$ in
the weakest fields $0.5\, H_{0},\, h/a=1.4$ case, and in the thicker
layer $H_{0},\, h/a=2$ case, whereas it should remain around $6\%$
for the thiner cells, higher field $H_{0}$ case. The extended potential
to take this lateral shift into account is also derived in section
\ref{ssect:gravit,effects}, and the dashed curve corresponds to this
potential at a fixed shift $z/h=14\%$, which is the maximum possible
in the thin cells, since it would correspond to contact of the particles
with one of the plates, coinciding with $z/h=(1-a/h)/2$ at $h/a=1.4$.
Considering this correction, one would expect the diamond and squares
to follow the dashed line, the circles to be close to the solid line,
and the triangle to sit in between. This is indeed approximately the
case at shorter times for the diamonds and rectangles, and the filled
circles fit well with the centered theory, both at short and long
times. Nonetheless, it is worth noticing that the open circles, collapse
rather with the lower field experiments than with the filled circles,
another experiment carried with identical parameters. This gives an
estimate of the reproducibility of these experiments, corresponding
roughly to a relative experimental error bar of $10\%$ for the separation
at a given time, estimated between open and filled circle cases. The
gravity-induced correction discussed above is of order $5\%$ for
the separation as function of the dimensionless time, and the effect
of this shift on the magnetic interactions can therefore hardly be
distinghished from experimental dispersion in terms of separation.
Still, the renormalization of the time due to hydrodynamic interactions
with the plates would lead in the thick cell case $h/a=2$, to $\eta /\eta _{0}\simeq 1.2$
if the particles were considered as centered $z/h=0$, instead of
the value $\eta /\eta _{0}=2.5$ that we used and corresponded to
the predicted shift in that case, $h/a=2,\, z/h=0.15$. This gravity-induced
correction is then clearly sensible for the hydrodynamic corrections,
if not so much on the magnetic interactions, since the above incorrect
viscosity $\eta $ would correspond to dividing $T$ by $2$,
which would double the abscissas of the square data points and drive
them way out of the theory and rest of the data. This shows qualitatively
that this shift was indeed present in cases $h/a=2,\, H_{0}$, and
$h/a=1.4,\, 0.5\, H_{0}$, and that particles sat close to or in contact
with a plate in this last case.

\paragraph{Corrections due to experimental error on the field directions.}
The fourth dotted curve represents the theoretical effect of a misalignment
of angle $\alpha =2.5^{\circ }$ between the constant field and the
normal direction. A direct generalization of sections \ref{sect:description}
to \ref{sect:isotropic} to such configurations with a slightly tilted
constant field, shows that the time averaged potential is still of
the form in Eqs. (\ref{eq:int,own,im,inst},\ref{eq:int,other,im,inst}),
with a modified parameter \begin{eqnarray}
\beta ' & = & \frac{\ah _{\perp }^{e}}{(1+\chi _{f})(\ah _{\parallel }+\sqrt{2}\sin (\alpha )\cos (\theta )\ah _{\perp }^{e})}\nonumber \\
 & \simeq  & \beta -\sqrt{2}(1+\chi _{f})\beta ^{2}\sin (\alpha )\cos (\theta ),\label{eq:modified,beta,for,tilted,stuff}
\end{eqnarray}
where $\theta $ is the angle between the projection of the constant
field over the plane and the separation vector. This modified interaction
potential leads to a torque tending to align the particle pair with
the direction $\theta =0$, and a radial interaction force corresponding
to a modified $\beta '$ in $[\beta -\sqrt{2}(1+\chi _{f})\beta ^{2}\sin (\alpha );\beta ]$,
$\beta '$ decreasing with time towards the lower limit as the
pair aligns with the inplane constant component of the field. The
dotted curve corresponds to this lower limit for a possible misalignment
$\alpha =2.5^{\circ }$, which is evaluated from the above as $\beta '=0.68$. 

For the longer time period shown in the encapsulated part of figure
\ref{Fig:adimen,time,sep,beta08}, an apparent equilibrium position
was reached in each experiment after typically $6\, T$ -- no more
than a 1\% relative motion was noticed later when the experiments
were several hours. This equilibrium separation corresponds
theoretically to the equilibrium one studied in the previous section,
$y_{s}(0.8)=2.35$. The high field and thin cell experiment (filled
circles) agree well (within 2\%) with the theory with a purely normal
field (solid line), but discrepancies between solid line and experiments
are noticeable at longer times for the four other experiments.
A comparison of the data with the dotted line shows that a misalignment
of order $2.5^{\circ }$ between constant field and normal direction
is sufficient to explain these discrepancies: in these experiments,
the particle pairs started at a relatively large angle $\theta $
from the inplane component of the constant field, which is why the
unmodified theory and experimental data are close for short times.
At longer times, the particle pairs aligned with the direction $\theta =0$,
and the modified theory $\beta '=0.68$ agrees well with the data.
An initial rotation of the particle pair and subsequent locking of
this direction in a particular one was indeed observed in these experiments. 

Brownian motion in the ferrofluid can be proved to be entirely negligible
for the relatively large particles and field we worked with, its relative
magnitude compared to magnetic interaction energy being $kT/\min \overline{U}\leq kT/[A(d^{3}/h^{3})\min u]\simeq 10^{-4}/\min u\leq 5\cdot 10^{-3}$
for the fields, particles and plate separations considered here.

\subsubsection{Scaled time-dependent results at various field geometries}

\begin{figure} {\par\centering \subfigure[h/d=1.4, with a potential $2.5^{\circ}$ misalignement between constant field and normal direction]{\resizebox*{0.45\textwidth}{0.30\textheight}{\includegraphics{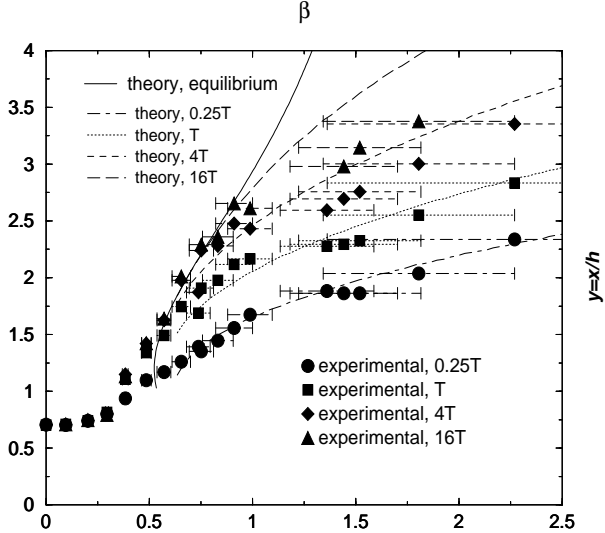}}} 

\subfigure[comparison between two plate separations]{\resizebox*{0.45\textwidth}{0.30\textheight}{\includegraphics{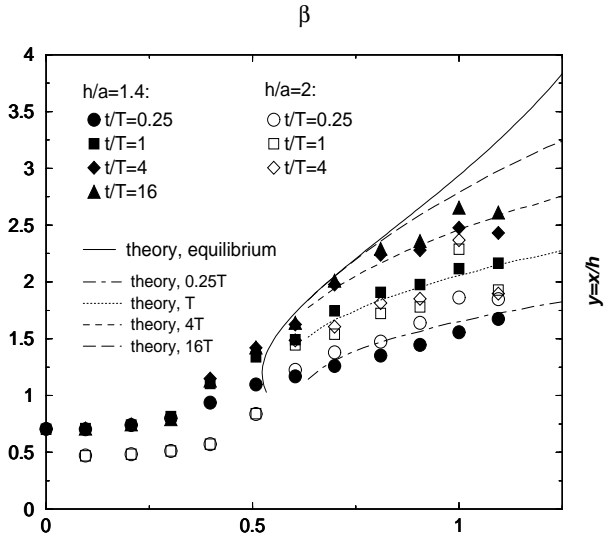}}} 

\par}

\caption{Separation as function of time and $\beta$ for various field configurations, times and cell sizes: merged data and theory.\label{Fig:merged}} \end{figure}

Experiments were carried out with inplane field magnitudes $H_{0}$,
and normal fields jumping at initial time from $0$ to $\beta (1+\chi _{f})H_{0}$
with various $\beta $ from 0 to 3.5. This was done for $a=50\, \mu m$
diameter particles, and plates separation $h=70\, \mu m$ and $100\, \mu m$.
The results, scaled separation as function of $\beta $, at four values of the scaled time, are shown in Figure \ref{Fig:merged}(a)
for the thinner cell. The error bars correspond to a possible misalignment
$\alpha =2.5^{\circ }$ between the constant field and the normal
direction: they represent the limits $[\beta -\sqrt{2}(1+\chi _{f})\beta ^{2}\sin (\alpha );\beta ]$
for the effective $\beta '$ parameter as explained in the previous
section. For the {}``c''-regime $\beta >\beta _{c}=1/\sqrt{2}\sim 0.7$,
theory and experiment agree well for any of the tested field parameters
and times. 

The solid line represents the theoretical equilibrium value, studied
in section \ref{sect:eq-sep-dist}. This is reached within typically
$16\, T$ (8 min for $h/a=1.4$, $H_{\parallel }=H_{0}$) for $\beta \leq 1$,
or longer time at higher $\beta $. This is the main reason why the
upwards curvature of the theoretical solid curve at larger values
of $\beta $ is not observed in experimental data, which correspond
to finite times, and for which other types of perturbations always
enter the picture at very large times and distances.

In Figure \ref{Fig:merged}(b), we present the results of experiments
carried out at two different plate separations, as function of time
and value of $\beta $. The error bars have been omitted for readability,
and the experimental points represented correspond to a constant field
supposed purely normal (i.e. the abcissas are the upper limit of the
error bars in Figure \ref{Fig:merged}(a)). The experiments carried
out in thicker cells, corresponding to weaker magnetic interaction
forces, are more sensitive any perturbations. The relative data collapse
for both plate separations at $0.25\, T$ and $T$ when $\beta >\beta _{c}$
nonetheless show that the separations in this regime scale with plate
separation, and not with particle diameter. Apart from the misalignment
of the constant field with the normal direction, a possible source
for these perturbations is as follows: when the particles come close
to equilibrium, inplane magnetic forces tend to zero, and the particle
motion becomes more sensitive to any interactions with the local environment
(confining plates) -- this being of course even more the case for
weaker fields or larger $h/a$: there seems to be a pinning (friction)
of the particles to an absolute plate position at large times. The
physical origin of this pinning is possibly due to roughness of the
plates (especially when particles are almost in contact), which can
quench the particles through the magnetic perturbation due to this
roughness (the repulsion effect of a dipole by its images would make
a particle sit preferably in positions of larger plate separation),
or alternatively when particles are almost in contact with the plates
can result in an inplane component of the hydrodynamic coupling or
contact forces responding to buoyancy forces. Instead of plate roughness,
the same type of qualitative effects could be due to small impurities
in the ferrofluids, starting to stick to the plates or particles at
large times, when the chemical surfactant layers around large particles
and possible impurities, break apart in some points.

For the regime $\beta <\beta _{c}$, the simple theory presented here
would predict that particles stay always in contact at $y=a/d=0.5$
or $0.71$ for $\beta <\beta _{m}\sim 0.55$, and for $\beta _{m}<\beta <\beta _{c}$
would either stay in contact if $a/h<y_{i}(\beta )$ or go to the
secondary minimum $y_{s}(\beta )$ if $y_{i}(\beta )<a/h$ (the separation
between first and second case happening at $\beta =0.61$ for $a/h=1/1.4,$
and $\beta =0.67$ for $a/h=1/2$). Particles seem indeed to be in
contact for $\beta \leq 0.2$, but start to separate significantly
well before $\beta _{m}$. We note also that this separation seems
grossly to be proportional to the particle diameter when $\beta \leq 0.3$,
where some finite separation can already be observed -- ordinates
of opened and filled symbols are multiple of each other through a
factor $100/70$ -- and proportional to plate separation in the regime
$\beta _{m}<\beta <\beta _{c}$. This shows that an extra physical
effect that was not taken into account here, generated repulsive forces,
whose range is finite but scales with the particle diameter. This
effect suppresses then the short range attraction in the regime $\beta _{m}<\beta <\beta _{c}$,
so that the particle jump directly to $y_{s}(\beta )$, which is always
a stable minimum and not a metastable one. For $\beta <\beta _{m}$,
this extra effect starts to separate the particles in proportion to
particle diameter. The physical origin of this short range repulsion
should not be the particle-particle hydrodynamic interactions, which
should slow down the relative motion rather than result in a net repulsion
-- see \cite{GB01,DSB+00} --. A probable candidate for this repulsion
is rather the magnetic effect of the finite size of the spheres, particularly
sensible when particles are close to contact: even if an isolated
nonmagnetic sphere generates a purely dipolar perturbation when it
is isolated in a homogeneous susceptible medium, this dipolar perturbation
does not fulfill the boundary conditions along the surface of another
magnetic sphere, sufficiently close of the first one to feel the heterogeneity
of the perturbation at the scale of its diameter, which is naturally
the case at a finite separation/diameter ratio. To model this short
range repulsion requires accounting for the magnetic perturbation
generated by this non pointlike character of close enough spherical
particles. Though this can be performed by a simple image method for
pairs of discs in 2D, one can show that this does not extend to pairs
of spheres in 3D, and the proper mathematical description of this
perturbation requires the use of a series of spherical harmonics, which
was not performed in the present study. To conclude this discussion,
we note that this correction seems to be negligible at separations
exceeding the sphere diameter $x/a>2$, as shows the agreement between
experimental data and the present theory when $\beta >\beta _{m}$.

Finally, we note that the present results are not contradictory
with similar experiments carried out in \cite{HS91} with a slightly
different ferrofluid, plate separation and particle size, where it
was reported that the equilibrium particle separation is approximately
linear in $\beta $ once particles have started to separate: for example,
this is also the case with the present ferrofluid, in the particular
case $h/a=1.4$ in the regime $0.3<\beta <1$ -- see the filled triangles
or theoretical curve in Figure \ref{Fig:merged}(a). This linear property
is however shown here to be a mere coincidence, for this does not
hold in the same $\beta $ regime for $h/a=2$, or for any $h/a$
when $\beta >1$, where $y_{s}(\beta )$ is curved upwards, and any
result $y(\beta ,t)$ observed at a given finite time $t$ is curved
downwards.

\section{generalization to quasi2D systems}

\label{sect:corrections}

\subsection{Gravity induced corrections}

\label{ssect:gravit,effects}

Although the effect of the dipolar images in the confining plates
tends to center particles at a mid-plane position, the density contrast
between the ferrofluid and particles tends to drive the particle out
of it for large enough particles. An estimation of this effect can
be obtained by considering for each particle, the sole effect of its
own images, plus buoyancy forces -- for the simple estimation we look
at here, we will neglect the coupling between one source and the images
of the other. Extending the analysis performed in sections \ref{sect:images}
and \ref{ssect:estab,isot} to a single dipole lying at a vertical
distance $z$ from the center between two horizontal plates, comes
directly\begin{eqnarray}
\overline{U}^{\text {shifted}}(z) & = & A\frac{a^{3}}{h^{3}}u^{\text {shifted}}(\frac{z}{h})\label{eq:U,shifted,with,dim}\\
u^{\text {shifted}}(s) & = & \sum _{l\in \mathbb{Z}^{*}}\kappa ^{\left|l\right|}\frac{1-2(-1)^{l}\beta ^{2}}{\left|l-(1+(-1)^{l+1})s\right|^{3}}\label{eq:U,shifted,adim}
\end{eqnarray}
This potential can be shown to be always centering for any value of
$\beta $, i.e. to have a single minimum in $s=0$, and to diverge
to infinity at $s=\pm 0.5$ -- plate contact for very small particles.
Equilibrium between gravity forces and magnetic interactions between
the dipole and its images leads to\begin{eqnarray}
\frac{d\overline{U}^{\text {shifted}}}{dz} & = & V(\rho _{g}-\rho _{f})\label{eq:equ,gravity,image,repuls}\\
\text {i.e.\, }\frac{du}{ds} & = & \frac{48(\rho _{g}-\rho _{f})gh}{\mu _{f}\bar{\chi }^{2}H_{\parallel }^{2}}\frac{h^{3}}{d^{3}}\label{eq:equ,gravity,repuls,nodim}
\end{eqnarray}
For small separations (i.e, small particles or strong enough fields),
a Taylor expansion to first order around the plates' center gives
\begin{eqnarray}
\frac{du}{ds} & = & 96\cdot C(\kappa )\cdot (1+2\beta ^{2})\cdot |s|\label{eq:im,repuls,taylor}\\
C(\kappa ) & = & \sum _{n\in N}\kappa ^{2n+1}/(2n+1)^{5}\label{eq:stupid,constant}
\end{eqnarray}
which is valid up to 25\% for $s<0.1$. Thus, the displacement looked
for can be evaluated as\begin{equation}
s=z/a=\frac{(\rho _{g}-\rho _{f})gh}{2C(\kappa )(1+2\beta ^{2})\mu _{f}\bar{\chi }^{2}H_{\parallel }^{2}}\frac{h^{3}}{a^{3}}\label{eq:equilibrium,shift}\end{equation}
when this quantity does not exceed $0.1$, or directly using Eqs.
(\ref{eq:equ,gravity,repuls,nodim},\ref{eq:im,repuls,taylor}) otherwise.
For the ferrofluid and particles we used, this led respectively for
$h=70\, \mu m,\, H_{\parallel }=H_{0};0.7\, H_{0}$ or $0.5\, H_{0}$,
and $h=100\, \mu m;\, H_{\parallel }=H_{0}$, to $s=0.06,\, 0.10,\, 0.15$
and $0.15$ .

The magnetic interaction potentials were unaffected up to 1\% in
the $s=0.06$ vertical shift case, and the dashed curve in Figure
\ref{Fig:adimen,time,sep,beta08} was obtained by considering particles
at a fixed $s=0.14$ out-of-midplane shift, using a generalized potential
obtained for a configuration sketched in Figure \ref{Fig:shifted,pair}
through an extension of the method used in sections \ref{sect:images}
and \ref{ssect:estab,isot}, \begin{figure} {\par\centering \resizebox*{0.35\textwidth}{0.07\textheight}{\includegraphics{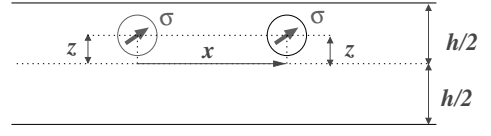}} \par}

\caption{Shifted hole pairs considered.\label{Fig:shifted,pair}} \end{figure} as \begin{eqnarray}
\lefteqn{u_{s}^{\text {pair}}(y)=(2\beta ^{2}-1)y^{-3}} &  & \label{eq:U,shifted,pair}\\
 &  & +4\sum _{l=1}^{+\infty }\kappa ^{l}\left(\frac{1+(-1)^{l}\beta ^{2}}{(y^{2}+m_{s}(l)^{2})^{3/2}}-\frac{3}{2}\frac{y^{2}+2(-1)^{l}m_{s}(l)^{2}\beta ^{2}}{(y^{2}+m_{s}(l)^{2})^{5/2}}\right)\nonumber \\
\lefteqn{m_{s}(l)=l+s\cdot \left[1+(-1)^{l}\right]} &  & \label{eq:auxiliary,shifted,pair}
\end{eqnarray}
This $s=0.14$ value was picked to represent the magnetic effect
of a shift sufficient to bring particles in contact with the plates
in the $h/a=1.4;\, 0.5\, H_{0}$ case. To an accuracy of 1\%, the
results for $s=0.15$ were very close to this case, the ones for $s=0.06$
very close to pure inplane situations, and the situation $s=0.10$
fell roughly halfway between both, and were therefore omitted from
Figure \ref{Fig:adimen,time,sep,beta08} for readability.

\subsection{Stability of the plane solutions and buckled configurations}

\label{sect:out-of-plane}

When particles are sufficiently small or fields sufficiently high,
the preceding section establishes that the confining plates have an
effective repulsive effect on an isolated particle, which is therefore
centered on midplane. In the case of a pair of particles, the interactions
between one source and the images of the other one might nonetheless
modify that picture, and make the plane solutions described in this
paper unstable as have been observed in some experiments. Neglecting
gravity, we will here generalize the interaction potential, to configurations
where particle pairs are allowed to tilt on both sides of the midplane,
i.e. where both are displaced by the same distance $z$ on both sides
of it -- cf Figure \ref{Fig:buckled,pair}. We consider only
symmetric situations, due to the symmetry of the problem under parity in the absence of gravitational forces.\begin{figure} {\par\centering \resizebox*{0.35\textwidth}{0.07\textheight}{\includegraphics{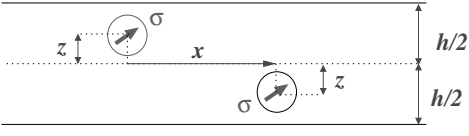}} \par}

\caption{Tilted hole pairs considered.\label{Fig:buckled,pair}} \end{figure} The effective interaction potential is obtained similarly to the
plane one, and comes as $\overline{U}(x,z)=A(a^{3}/h^{3})\cdot u(x/h,z/h)$
where\begin{eqnarray}
u(y,s)\! \!  & \! \! =\! \!  & \! \! 2\! \sum _{l=-\infty }^{+\infty }\! \kappa ^{|l|}\! \left(\frac{1+(-1)^{l}\beta ^{2}}{(y^{2}+(l+p_{l}s))^{2})^{3/2}}-\right.\label{eq:U,tilted}\\
 &  & \qquad \qquad \quad \left.\frac{3}{2}\frac{y^{2}+2(-1)^{l}(l+p_{l}s)^{2}\beta ^{2}}{(y^{2}+(l+p_{l}s)^{2})^{5/2}}\right)\nonumber \\
 & \! \! \! \!  & \! \! +2\! \sum _{l\in \mathbb{Z}^{*}}\! \kappa ^{|l|}\! (1\! -\! 2(-1)^{l}\! \beta ^{2})\! \left(\frac{1}{|l+(2-p_{l})s|^{3}}\! -\! \frac{1}{|l|^{3}}\! \! \right)\nonumber \\
p_{l} & = & 1+(-1)^{l}\label{eq:parity,function}
\end{eqnarray}
which reduces to the previous in-plane solution, Eq. (\ref{eq:U,av,im,isot}),
when $s=0$. Contour plots of this pair potential are displayed in
Figure \ref{Fig:buckled,pot}, for the ferrofluid used here, $\chi _{f}=1.9$,
and the four types of potentials -- $\beta $ values are identical
to the ones adopted in section \ref{ssect:propert,isot}. Black and
white represent respectively $u=-0.2$ and $0.5$ in subfigures (a)
and (b), $u=-0.2$ and $1$ in (c) and (d), and the grey level linear
in $u$ in between. Out-of-plane values represented cover the whole
possible range $0\leq s\leq 0.5$ in (c) and (d), and are restricted
to 10\% from the midplane in (a) and (b). \begin{figure} {\par\centering \subfigure[$\beta<\beta_m$]{\resizebox*{0.23\textwidth}{0.17\textheight}{\includegraphics{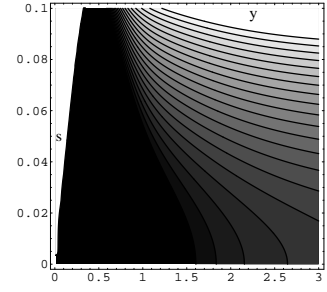}}} \par}

{\par\centering

\subfigure[$\beta_m<\beta<\beta_c$]{\resizebox*{0.23\textwidth}{0.17\textheight}{\includegraphics{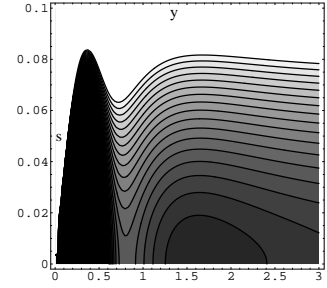}}}\par}

{\par\centering \subfigure[$\beta_c<\beta<\beta_u$]{\resizebox*{0.23\textwidth}{0.17\textheight}{\includegraphics{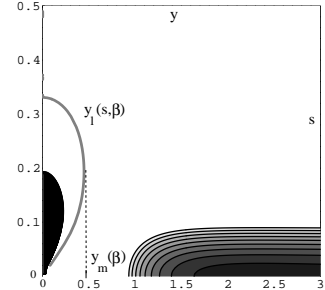}}}\par}

{\par\centering

\subfigure[$\beta_u<\beta$]{\resizebox*{0.23\textwidth}{0.17\textheight}{\includegraphics{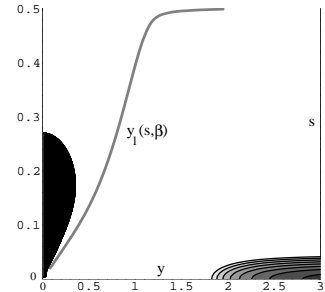}}} \par}

\caption{Contour plots of the interaction potential as function of inplane separation y and tilt coordinate s.\label{Fig:buckled,pot}} \end{figure} The inplane configurations correspond to the bottom axis of those
graphs. 

Inplane solutions are in principle locally stable if $\partial ^{2}u/\partial s^{2}(y,0)>0$,
otherwise particle pairs will tend to tilt. For both first cases,
$\beta <\beta _{c}$, we note that $\partial ^{2}u/\partial s^{2}>0$
for any possible $(y,s)$, and any tilt is restored by the magnetic
image effect: the plane configurations are indeed stable. As soon
as $\beta >\beta _{c}$, besides the minima $(y,s)=(y_{s},0)$ or
$(+\infty ,0)$ in c or d case, another local minimum appears at a
certain $(0,s_{e}(\beta ))$: particles can be stable at a finite
distance on top of each other -- the pair tends to align with the
large normal field --, or if they are close enough to be attracted
by this potential minimum but too large ($a>s_{e}(\beta )$ and $a>h/2$),
contact forces between them and with the confining plates will attract
them to a {}``buckled'' configuration where both particles are in
contact with each other, and with one different plate each. Note that
$0<s_{e}(\beta )<0.5$, i.e. very small particles $a\ll h$ at this
second minimum would sit on top of each other, neither in contact
between them nor with the plates. The criterion to determine whether
particles are attracted by the inplane solution, or the buckled configurations,
is to determine whether the present $(y,s)$ lie in the basin of attraction
of one minimum or the other. Both basins of attraction are separated
by a ridge of the potential, on which $u$ decreases under the effect
of any perturbation of $(y,s)$ apart from the ones directed exactly
along its gradient. This boundary between both basins of attraction,
noted $y_{l}(s,\beta )$ was determined numerically and plotted as
the grey solid line in Figure \ref{Fig:buckled,pot} (c) and (d).
We note that the whole axis $(y,0)$ lies in the basin of attraction
of the plane minimum, so any particle pair starting with no tilt should
end up so. However, at short enough distances $y$ in a c-field, or
at any inplane distance in a d-field, certain configurations with
a finite tilt are attracted by the normal-aligned pair minimum, and
will end up in a buckled contact configuration or normal-aligned pair.
We have determined for every $\beta $, the maximum $y_{m}(\beta )=\max _{s}y_{l}(s,\beta )$:
for a given $\beta $, when $y>y_{m}(\beta )$ any tilted configuration
will be attracted by the inplane solution, whereas for certain finite
tilts at close enough $y<y_{m}(\beta ),$ the pairs will be attracted
towards buckled in-contact configurations. The function $y_{m}(\beta )$
is displayed as the dashed curve in Figure \ref{Fig:stab,lim,x,beta}
-- the solid curve in a reminder of the equilibrium inplane distances
$y_{s}(\beta ),y_{i}(\beta )$ determined in section \ref{sect:eq-sep-dist}.
The function $y_{m}(\beta )$ diverges when $\beta \rightarrow \beta_u^{-}$,
illustrating the fact that in any d-type field, configurations with
out-of-plane tilts $s$ close to $0.5$, i.e. both particle centers
almost along the plates (for small enough particles) will be attracted
by the normal-aligned pair mode.\begin{figure} {\par\centering \resizebox*{0.30\textwidth}{0.25\textheight}{\includegraphics{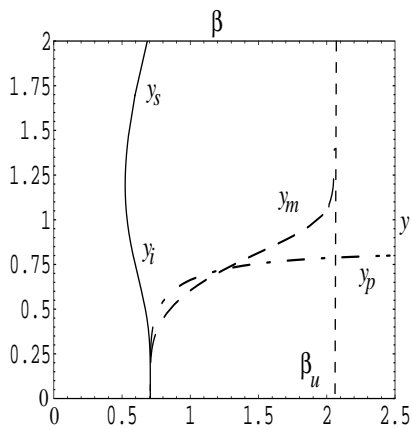}} \par}

\caption{Stability limits of the in-plane configurations. \label{Fig:stab,lim,x,beta}} \end{figure}

This effect is believed to be responsible for the lattices of buckled
chains of particles in contact observed in \cite{SH91}. These were
observed in pure normal fields ($\beta =+\infty )$, and the non-trivial
character of the lattices, being hexagonal or square instead of triangular
lattices characteristic of attractive interactions at any range, can
be qualitatively explained by frustration effects: neighboring particles
tend to sit in top-down contact configurations, but top-top or down-down
configurations are repulsive -- cf shifted potential developed in
the previous section -- and therefore lattices with noneven number
of particles along the loops, as the triangular one, are disfavored
in comparison with those involving even numbers in loops, as the square
or hexagonal ones. 

Eventually, the local stability of a plane configuration was investigated
in fields of c- or d-type: the second derivative $\partial ^{2}u/\partial s^{2}$
is positive at $s=0$ for large enough distances, and small out-of-plane
tilts will be restored by magnetic interactions, but below a certain
$y<y_{p}(\beta )$, $\partial ^{2}u/\partial s^{2}(y,s=0)<0$, and
the inplane solution will be locally instable. However, the preceding
established that this out-of-plane character will be transient, since
this case will still be attracted at long times by the inplane solution.
This upper limit $y_{p}(\beta )$ below which inplane configurations
will go to a transient out-of-plane regime, was determined numerically
for any $\beta $ and plotted as the dash-dot curve in Figure \ref{Fig:stab,lim,x,beta}.

\section{Conclusions}

For pairs of magnetic holes in ferrofluid layers of finite thickness,
exposed to fast oscillating conic magnetic fields, we have established
the effective interaction potential driving their slow motion. The
importance of the magnetic permeability contrast between ferrofluid
and confining plates on these interaction was demonstrated, and the
resulting pair potential analytically derived. This allowed the classification of
those interaction potentials into four types, two of which present a
secondary minimum at a finite distance. A simple finite time theory
for these non-Brownian microspheres, including the hydrodynamic
interactions of the holes with the confining plates, was directly
compared and confirmed by experimental results, through data collapse
of the scaled separation as function of scaled time, for various plate
separations and field magnitudes. The relaxation time of these systems
to reach equilibrium when there is any, was typically a few minutes
or above. Eventually, we generalized the study to full three dimensions
in the layer thickness, and studied the stability of the in-plane
configurations. This established that although inplane configurations
are stable at small normal fields, or at large ones and sufficient
particle separation, hole pairs can be attracted by another stable
configuration of tilted pairs of particles with contact between them
and contact with one plate each.

This simple theory renders for so far unexplained observed configurations
of magnetic holes, namely 2D lattices with finite separation, or buckled
lattices of tilted pairs of particles. In principle, the effect described
here in detail should be important in any colloidal system confined
in a layer with significant magnetic permeability or dielectric contrast
between the fluid and confining structure.

The ability to tune the equilibrium distance at will through the ratio
of the normal over in-plane magnitude of the external field, makes
this magnetic hole system a good candidate for various applications,
as the manipulation of large molecules using the magnetic holes, for example proteins which would be fixed to one or several holes coated with antigens, the
determination of the transport properties of a ferrofluid, or as analog
model of systems implying discrete particles of tunable interactions
and hydrodynamic coupling to the carrier fluid. The effective pair
interactions derived here should be a fundamental brick of any such
applications.

\bibliographystyle{apsrev}

\end{document}